\documentclass[twocolumn,prl,showpacs,preprintnumbers,amsmath,amssymb]{revtex4}
\usepackage{graphicx}
\usepackage{epsfig}
\usepackage{dcolumn}
\usepackage{bm}

\begin{document}

\title{Electron Transport in a Multi-Channel One-Dimensional Conductor: Molybdenum Selenide Nanowires}

\author{Latha Venkataraman}
\author{Yeon Suk Hong}
\author{Philip Kim}
\affiliation{Department of Physics, Columbia University, New York,
New York 10027}

\date{\today}

\begin{abstract}

We have measured electron transport in small bundles of identical
conducting Molybdenum Selenide nanowires where the number of
weakly interacting one-dimensional chains ranges from 1-300. The
linear conductance and current in these nanowires exhibit a
power-law dependence on temperature and bias voltage respectively.
The exponents governing these power laws decrease as the number of
conducting channels increase. These exponents can be related to
the electron-electron interaction parameter for transport in
multi-channel 1-D systems with a few defects.

\end{abstract}

\maketitle

Interacting electrons in one-dimensional (1D) metals constitute a
Luttinger Liquid (LL)\cite{Voit}, in contrast to a Fermi liquid
(FL) in 3-dimensional (3D) metals. Transport properties of 1D
conductors are strongly modified as adding an electron to a 1D
metal requires changing the many-body state of its collective
excitations. This results in vanishing electron tunneling density
of states at low energy. Power-law dependent suppression in
tunneling conductance has been observed in many systems, including
fractional quantum Hall edge states \cite{Chang}, single and
multi-walled carbon nanotubes \cite{Bockrath,Yao,Bachtold},
bundles of NbSe$_3$ nanowires \cite{Slot} and conducting polymers
\cite{Aleshin}. A cross-over from a truly 1D Luttinger-liquid (LL)
to a 3D Fermi-liquid (FL) is expected as 1D conductors are coupled
together, increasing the number of weakly interacting
channels~\cite{Matveev, Sandler}. This transition, however, has
not been observed in the above (quasi) 1-D systems due to the
experimental difficulty in preparing identical conducting quantum
wires to form conductors with a few weakly interacting channels.
In this letter, we report temperature and bias dependent electric
transport measurements on small bundles of Molybdenum Selenide
(MoSe) nanowires~\cite{Tarascon1,Golden,Venkataraman}, whose
diameter ranges from 1-15~nm. These nanowires, which consist of
bundles of weakly interacting and electrically identical 1D MoSe
molecular chains, show a power-law dependent tunneling
conductance. The exponent governing the power-law decreases as the
bundle diameter increases, indicating a transition from 1D to bulk
transport with increasing number of conducting channels.

\begin{figure}
\epsfig{file=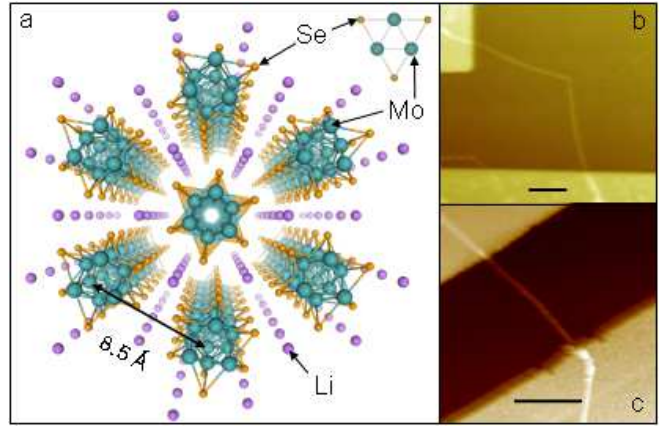,width=\linewidth}
\caption{\label{fig:AFM}(color online) (a) Structural model of a
7-chain MoSe nanowire along with the triangular Mo$_3$Se$_3$ unit
cell, (b) and (c) AFM height images of MoSe nanowires between two
Au electrodes. The wire heights are 7.2 nm and 12.0 nm
respectively. Scale bar $=$ 500 nm.}
\end{figure}

Crystalline bundles of MoSe chains are obtained from the
dissolution of quasi-1D Li$_2$Mo$_6$Se$_6$ crystals in polar
solvents. Single crystal Li$_2$Mo$_6$Se$_6$ was prepared as
described previously \cite{Tarascon2}. X-ray diffraction analysis
showed hexagonally close packed molecular MoSe chains with a
lattice spacing $a_0  = 0.85$~nm, separated by Li atoms
(Fig.~\ref{fig:AFM}(a)). Atomic scale bundles of MoSe nanowires
were produced from $\sim$100~$\mu$M solutions of
Li$_2$Mo$_6$Se$_6$ in anhydrous methanol. The solutions were then
spun onto degenerately doped Si/SiO$_2$ substrates with
lithographically patterned electrodes in a Nitrogen atmosphere.
Typically, 35~nm thick Au electrodes with 5~nm Cr adhesion layer
separated by $\sim 1~\mu m$ were used to contact randomly
deposited nanowires. Figs.~\ref{fig:AFM}(b) and (c) show atomic
force microscope (AFM) images of typical devices. The two-probe
resistance of such a device, which ranged from $\sim$~100~k$\Omega
- $100~M$\Omega$ at room temperature, was measured in a cryostat
with a continuous flow of helium. A degenerately doped silicon
substrate, underneath a $L_{ox}=300$~nm thick silicon dioxide
dielectric layer, served as a back gate to modulate the charge
density in the nanowires. Once transport measurements were
complete, the wire diameter, $D$, was determined from an AFM
height profile.

%~\footnote{Although these wires tend to oxidized in that they
%become insulating after prolonged exposure to air, their diameter
%does not change}.

\begin{figure}
\epsfig{file=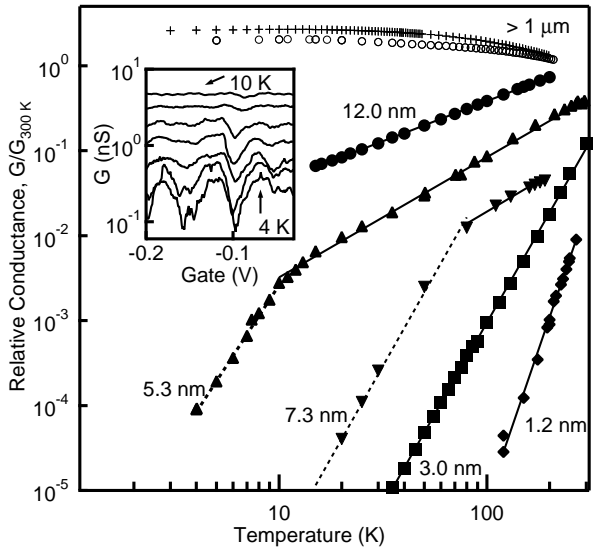, bb = 0 0 314 250,width=\linewidth}
\caption{\label{fig:GvsT}  Relative conductance ($G/G_{300 K}$)
versus $T$ for six mesoscopic and two bulk MoSe wires labelled
with the wire diameter. Curves are offset vertically for clarity.
The solid lines are power-law fits to the data ($G \sim
T^\alpha$). The dashed line indicates the region where Coulomb
blockade becomes important in two wires ($\blacktriangle$ and
$\blacktriangledown$). Inset: Gate dependence of conductance (G)
for the 5.3 nm wire ($\blacktriangle$) from 4K to 10 K at 1K
intervals.}
\end{figure}

Fig.~\ref{fig:GvsT} shows the conductance $(G)$ normalized by its
room temperature value as a function of temperature $(T)$ for a
representative subset of the samples studied \footnote{The
mesoscopic wires were susceptible to structural deformations
locally at the wire/electrode junctions. Due to the invasiveness
of the electrodes, multi-terminal measurements were not
possible.}. We applied a small bias voltage, $(V \ll k_BT/e)$, to
stay in the linear response regime for this measurement. Notably,
the mesoscopic scale samples ($D<20~$nm, $L\sim 1~\mu$m) exhibit
more than two orders of magnitude conductance decrease with
decreasing temperature in the measured temperature range, unlike
the samples in the bulk limit ($D>1~\mu$m, $L\sim100~\mu$m), which
exhibit a bulk metallic behavior, as the conductance increases
with decreasing temperature, i.e. energy. A power-law dependence,
$G \sim T^\alpha$, is evident in these mesoscopic samples, where
the exponent $\alpha$ can be readily extracted from the slope of
the least-squares fit line in the double logarithm plot. For most
of the samples, $G(T)$ can be expressed by a single $\alpha$
within the experimentally accessible conductance
range~\footnote{Since $G$ decreases rapidly as $T$ decreases in
the nanowire samples, most of our $G(T)$ data were limited at low
$T$ by our current sensitivity ($10^{-13}A$).}. However, for some
wires with a relatively high conductance ($>1~\mu$S) at room
temperature (Fig.~\ref{fig:GvsT} $\blacktriangle$ and
$\blacktriangledown$), an abrupt change in the exponent at low
temperatures was observed. In this low temperature regime, the
conductance varied with the gate voltage (Fig.~\ref{fig:GvsT}
inset) and the exponent depended on the applied gate voltage,
$V_g$, due to Coulomb charging effects. Above this Coulomb
charging temperature, a general trend of decreasing $\alpha$ with
increasing $D$ is found for all mesoscopic samples studied. This
trend will be discussed further later in the paper.
%In order to rule out the possibility that our measurements are
%caused by the contacts to the wires, we point out that the room
%temperature two probe resistance varied between
%$\sim$~100~k$\Omega$ and $\sim$~100~M$\Omega$ from wire to wire
%and was not correlated to the exponent $\alpha$. In addition for
%one wire measured which spanned 3 electrodes, the two probe
%resistance of the two segments were different by a factor of 2,
%indicating that the contacts to the two segments were indeed
%different, however, the exponent from temperature dependent
%transport measurements were identical\footnote{This wire was not
%listed in Table~\ref{tab1} as it was destroyed during the
%measurements}.

We now consider several possible explanations for seeing a
decreasing conductance with decreasing temperature. For example,
one expects $ln(G) \propto -1/T$ for barrier activated transport.
For highly defective wires, one expects $ln(G) \propto -
1/T^{\delta}$ due to variable range hopping between localized
states in the wire, where $\delta$ can range from $1/4$ for a 3D
wire to $1/2$ for a 1D wire \cite{Imry}. However, neither of these
models fit our data as, for all mesoscopic scale measured
irrespective of wire diameter, the conductance follows a power-law
remarkably well. Another potential explanation is a non-metallic
behavior associated with a Peierls transition, which opens up an
energy gap at the Fermi level. However, from the conduction
measurements on the bulk quasi-1D crystals ($D > 1 \mu$m) and also
from previous scanning tunneling microscopy work on similar
nanowires~\cite{Venkataraman}, we observe no evidence for a gap
opening at temperatures down to 5~K, consistent with band
calculations \cite{Ribeiro}. Alternatively, a power law dependent
tunneling conductance is predicted for tunneling into a Luttinger
Liquid, for 1D Wigner Crystals \cite{Maurey}, for a highly
disordered systems where the electron mean free path is comparable
with the wire diameter \cite{Egger}. We can rule out a 1D Wigner
Crystal model since the Coulomb interactions, which are screened
by the back gate, are not long-ranged. We also eliminate the
scenario for a strongly disordered system by estimating the
electron mean free path, $l_e$ in the nanowire. We can estimate
$l_e$ indirectly from the effective wire length, $L_{eff}$,
obtained from the dependence of the Coulomb charging energy on the
gate voltage. Here, the charging energy $E_c = e^2/C$, where the
wire capacitance $C \approx 2\pi\epsilon L_{eff}/ln(4L_{ox}/D)$.
For the nanowire device shown in the inset of Fig.~\ref{fig:GvsT},
the estimated charging energy $E_c$ is 5-10~meV from the
conductance map in $V$ and $V_{g}$ (not shown), from which we
obtained $L_{eff} \sim 0.3-0.6~\mu$m. With this estimate, we rule
out the possibility of having a highly disordered system.
Moreover, the fact that the measured resistance is larger than
$\sim$100~k$\Omega$ and is not directly correlated with the wire
diameter indicates that transport is dominated by tunnelling. Thus
a model concerning tunnelling into a relatively clean LL is a more
likely description of the observed transport phenomena
 \footnote{We note that the environmental Coulomb blockade (ECB)
model (see for example, Ingold and Nazarov, cond-mat/0508728)
might be applied to our experiment. However, its validity is
limited to the large N limit~\cite{Matveev}, where the LL and ECB
models do not offer any observable differences in our experimental
setup and we thus consider only the LL model in this paper.}.

\begin{figure}[b]
\epsfig{file=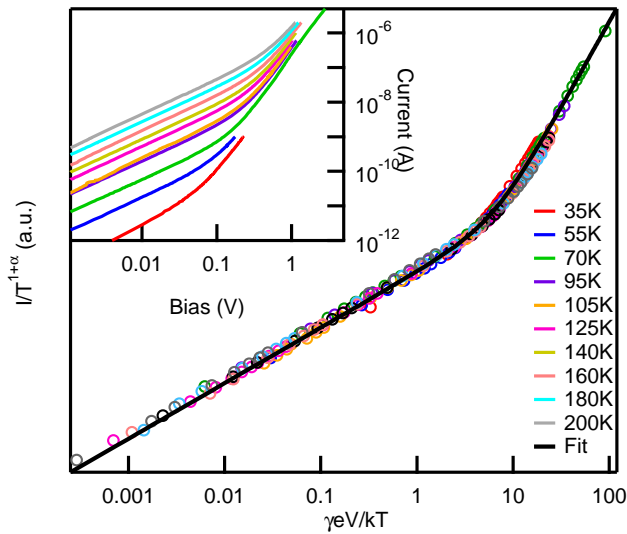, bb = 0 0 307 255,width=\linewidth}
\caption{\label{fig:IvsV}(color) Inset: Wire W3 I-V data taken at
different temperatures between 35 K and 200 K. All curves show a
change from linear response to power-law dependence at a
temperature dependent bias voltage. Main panel: $I/T^{\alpha+1}$
determined from I-V data plotted against $\gamma eV/k_BT$ with
Eq.~\ref{IVFit} fit to the data. The measured exponents are
$\alpha = 4.3$ and $\beta = 2.1$ ($\sim \alpha/2$) and the fitting
parameter $\gamma$ is $0.25 \pm 0.1$.}
\end{figure}

\squeezetable
\begin{table*}
\caption{\label{tab1}Measured exponents $\alpha$ and $\beta$
determined from the temperature and bias dependent conductance
measurement, along with the wire diameter (D) as determined by AFM
and the number of channels including spin (N) calculated from D.
Wires indicated by asterisk (*) have $\alpha \approx \beta$ but
for all other wires, $\alpha \simeq 2\beta$.}
\begin{ruledtabular}
\begin{tabular}{c|c|c|c|c|c|c|c|c|c|c|c|c|c}
  Wire & W1 &W2* &W3 &W4 &W5 &W6 &W7 &W8* &W9 &W10 &W11 &W12* &W13 \\ \hline
 $\alpha$&6.6&5.2& 4.3& 3.95&2.33& 1.40& 2.34& 1.1& 1.55& 1.95& 1.2& 0.94& 0.61\\
 $\beta$&3.0&4.9 &2.1 &1.9  &1.0& 0.72&1.2& 1.0& 0.8&1.09& 0.6& 0.90& 0.32\\
 $D$&$0.8\pm0.5$&$ 2.1\pm0.3 $&$3.0\pm0.3 $&$3.5\pm0.2 $&$ 5.0\pm0.5$
&$ 5.3\pm1.0$&$6.1\pm0.5$&$7.2\pm0.5$&$7.3\pm1.4 $&$7.4\pm1.5 $&$10.3\pm0.4 $&$12.0\pm0.7 $&$15.7\pm1.1 $\\
 $N$&2&12&22 &30  &62&70&94&130&134&138&268&362&620\\
\end{tabular}
\end{ruledtabular}
\end{table*}

Further support for the LL-like transport in the MoSe nanowires
can be found in the bias dependence of the conductance in the
non-linear response regime. According to the LL model in a
tunneling regime~\cite{Voit}, the bias voltage dependent transport
current, $I(V)$, has a transition between an Ohmic behavior, i.e.
$I\propto V$ in the low bias regime ($V \ll k_BT/e$), and a power
law behavior with an exponent $\beta$, i.e., $I \propto
V^{\beta+1}$ in the high bias regime ($V \gg k_BT/e$). The inset
of Fig.~3 shows typical $I(V)$ data measured in a mesoscopic wire
with the applied bias voltage ranging over more than three orders
of magnitude at different temperatures. A transition between Ohmic
and power-law behavior is observed as $V$ increases.

Interestingly, we also found that the exponent $\beta$ depended
strongly on $D$, as can be seen in Table ~\ref{tab1}, where we
list $\alpha$, $\beta$ and $D$ for 13 samples. In general, we find
that $\alpha$ decreases monotonically as $D$ increases. Based on
the relation between $\alpha$ and $\beta$, we can categorize our
samples into two distinct groups: group (I) where $\alpha \approx
2 \beta$; and group (II) where $\alpha \approx \beta$. In our
experiments, the majority of samples (10 out of 13 in
Table~\ref{tab1}) belongs to group (I), while only a few samples
(3 out of 13 in Table~\ref{tab1}) belong to group (II)
\footnote{Group (II) samples can be sub-divided further into a
group with notable defects such as the one shown in
Fig.~\ref{fig:AFM}(b) ($\gamma \approx$1) and a group without
strong defects ($\gamma \approx$1/2), where the parameter $\gamma$
is defined in Eq.~\ref{IVFit}.}.

For a clean LL without defects, these two exponents are expected
to be identical, i.e., $\alpha = \beta$~\cite{Voit}, since they
are characteristic of a single junction between FL and LL. The
deviation from this model found in group (I) samples can be
explained within the LL model with a few defects as described
below. Strong defects in the nanowires break the conducting
channels into a few LL dots connected in series between the
electrodes. In this multiple LL dot scheme, the wires have two
kinds of tunnel junctions; (i) junctions between the electrode and
wire, constituting a FL to the end of LL junction; and (ii)
junctions between two wire segments, constituting an LL to LL
junction. In such a wire, the tunneling probability can be
specified by two distinct exponents $\alpha_{LL-LL}$ and
$\alpha_{FL-LL}$, where $\alpha_{LL-LL}=2\alpha_{FL-LL}$
holds~\cite{Bockrath}. If the linear response resistances of the
FL-LL and LL-LL junctions at room temperature are of similar
orders of magnitude, we expect $G \propto T^{\alpha_{LL-LL}}$ in
the low temperature limit, since the LL-LL junctions become most
resistive. However, in the high bias limit the FL-LL junctions
become more resistive than the LL-LL junctions since
$\alpha_{LL-LL}>\alpha_{FL-LL}$, and thus $I \propto
V^{\alpha_{FL-LL}+1}$. Therefore the exponents obtained from
temperature and bias dependent data are expected to have a
relation $\alpha = 2\beta$ for wires with a few defects that break
them into multiple LL dots, as observed in our group (I) samples
\footnote{For temperatures below 300 K, the deviation of the
measured $\alpha$ from $\alpha_{LL-LL}$ is less than 10\% as long
as $R_{Lead-Wire}/R_{Defect}$ is between $\sim 0.1 - 10$ where
$R_{Lead-Wire}$ is the total resistance between the wire and the
Au lead and $R_{Defect}$ is the total resistance of all defects. A
similar tolerance range for $R_{Lead-Wire}/R_{Defect}$ is also
found for $\beta \approx \alpha_{FL-LL}$. We have observed a few
devices outside of this tolerance limit, which shows an
intermediate bias ranges where LL-LL tunnelling is dominant at low
temperature.}. This argument does not hold however if there are no
defects in the wire or in the extreme of strong defects that
dominate transport within the experimentally accessible range of
$T$ and $V$. For such samples, $\alpha \approx \beta$, which is
the relation we find in group (II) samples.

The power-law behaviors in $T$ and $V$ allow us to scale $I(V,T)$
into a single curve~\cite{Bockrath,Balents}. Considering the above
arguments, we can modify the scaling formula of a clean LL
transport model to include the two exponents $\alpha$ and $\beta$
as:
\begin{equation}
 I = I_0 T^{\alpha+1} \sinh\left(\frac{\gamma eV}{2k_BT}\right)
 \left|\Gamma\left(1 + \frac{\beta}{2}+\imath \frac{\gamma eV}{2k_BT}
\right)\right|^2\label{IVFit}
\end{equation}
where $I_0$ and $\gamma$ are constants independent of $T$ and $V$.
Physically, $\gamma$ represent the ratio between the voltage
across a dominantly resistive junction at high bias to the applied
bias voltage~\cite{Bockrath}. As shown in Fig.~\ref{fig:IvsV}, the
series of $I(V)$ curves measured at different temperatures for the
same sample collapse remarkably well onto a single curve described
by Eq.~\ref{IVFit} over the entire measured temperature range by
plotting $I/T^{\alpha+1}$ against $\gamma eV/k_BT$ with only one
fitting parameter $\gamma$. For the data in Fig.~\ref{fig:IvsV},
$\gamma$ is $0.25\pm0.1$, implying that there are probably four
barriers of approximately equal resistance over which the applied
bias voltage is distributed.

\begin{figure}
\epsfig{file=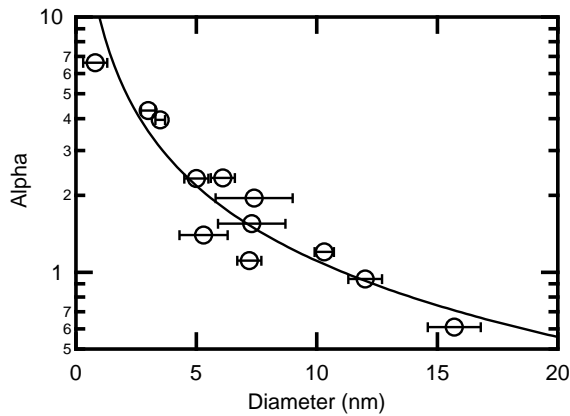, bb = 0 0 270 174,width=\linewidth}
\caption{\label{fig:AlphavsD} The exponent $\alpha$, plotted
against the wire diameter $D$. The solid line is a fit to
Eq.~\ref{GvsN}.}
\end{figure}

Finally, we now discuss the dependence of the exponents on the
wire diameter ($D$) in order to elucidate the transition from a
few channel 1D transport to the 3D transport limit. For this
purpose, we focus on the samples with $\alpha = \alpha_{LL-LL}$
(i.e. group (I) samples and the group (II) samples with $\gamma
\approx$~1). In Fig.~\ref{fig:AlphavsD}, we show the measured
$\alpha$ plotted against $D$. Since $N \propto D^2$, where $N$ is
the number of channels in the wire including the spin degree of
freedom~\footnote{We use $N = 2f D^2/a_0^2$ where $f$, the filling
factor, is 0.91 assuming a hexagonal close packing of the
chains.}, the observed rapid decrease of $\alpha$ for large
diameter nanowire bundles indicates a cross over from 1D behavior
to 3D transport ($\alpha \approx 0$). Employing the
electron-electron interaction parameter for a single chain (N=2),
$g$, the end tunneling exponent $\alpha_{LL-LL}$ for an $N$
channel LL wire can be expressed as~\cite{Matveev, Egger2}:
\begin{equation}
\alpha_{LL-LL} =
\frac{2}{N}\left[\left(1+NU\right)^{1/2}-1\right], \label{GvsN}
\end{equation}
where $U\simeq 2/g^2$. We fit this equation to our data using g as
a single fitting parameter. A good agreement with the experimental
observation was obtained for $g=0.15$. For a screened Coulomb
interaction, $g$ can be estimated by $g \simeq
\sqrt{1/(e^2ln(4L_{ox}/a_0)/\pi\hbar v_F \kappa)}$~\cite{Matveev},
where $v_F$ is the Fermi velocity of a single chain and $\kappa$
is the dielectric constant of silicon dioxide. From the fit in
Fig.~\ref{fig:AlphavsD}, we deduce that $v_F = 3\times 10^4$m/s.
We note here that this value is smaller than the value obtained
from recent band calculation ($4 \times 10^5$m/s)~\cite{Ribeiro},
indicating that a static screening picture considered in this
model might be too simplistic. Further theoretical considerations
including the effect of impurities and inter-chain hopping are
needed to elucidate strongly interacting electrons in these 1D
channels.

%According to a theoretical model for transport in multi-channel
%wires \cite{Matveev, Egger2}, we should find, in the limit of
%large N,
%\begin{equation}
%\alpha \simeq \sqrt{V_0/\pi\hbar v_F N}. \label{GvsN}
%\end{equation}
%Here $V_0 = 2e^2 ln(4L_{ox}/a_0)/\kappa$ is the Coulomb potential
%screened by the gate, $a_0$ is the diameter of a single chain and
%$\kappa$ is the dielectric constant of silicon dioxide. In
%Fig.~\ref{fig:AlphavsD} we plot our measured $\alpha$ values
%against $D$, and fit Eq.~\ref{GvsN} to this data. Within an
%LL-picture, the interaction parameter for a single wire determined
%from the fit since is $g_1 = (1+V_0/\pi\hbar v_F)^{-1/2} = 0.08
%\pm 0.01$(CHECK EQ.).

%In conclusion, we have presented measurements of electron
%transport in mesoscopic MoSe nanowire samples over a wide range of
%temperatures and bias voltages. Prominent power law behaviors in
%transport support that these wires constitute a Luttinger liquid
%characterized by strongly repulsive electron-electron
%interactions. In addition, a clear trend from 1D transport to 3D
%transport is observed as the number of conduction channel
%increased in the wires.

We thank C. M. Lieber, Y. Oreg, A. Millis, I. Aleiner, B.
Altshuler and R. Egger useful discussions. This work was supported
by NSF Award Number CHE-0117752, by the New York State Office of
Science, Technology, and Academic Research (NYSTAR). This work
used the shared experimental facilities supported by MRSEC Program
of the NSF (DMR-02-13574). Y.S.H acknowledges support from the
Korean Science and Engineering Foundation. P.K. acknowledges
support from NSF CAREER (DMR-0349232) and DARPA
(N00014-04-1-0591).

\bibliography{TransportPaper}

\end{document}